\documentclass[preprint,eqsecnum,aps,nofootinbib]{revtex4}
\usepackage{amsfonts,amsmath,amssymb,amsthm}
\usepackage{latexsym}
\usepackage{bbm,bm}
\usepackage{graphicx}

\newcommand{\beq}{\begin{equation}}
\newcommand{\eeq}{\end{equation}}
\newcommand{\beqs}{\begin{eqnarray}}
\newcommand{\eeqs}{\end{eqnarray}}
\begin{document}

\title{Dynamics of Entanglement and Uncertainty Relation in Coupled Harmonic Oscillator System : Exact Results}

\author{DaeKil Park$^{1,2}$}

\affiliation{$^1$Department of Electronic Engineering, Kyungnam University, Changwon
                 631-701, Korea    \\
             $^2$Department of Physics, Kyungnam University, Changwon
                  631-701, Korea    
                      }

\begin{abstract}
The dynamics of entanglement and uncertainty relation is explored by solving the  time-dependent Schr\"{o}dinger equation for coupled harmonic
oscillator system analytically
when the angular frequencies and coupling constant  are arbitrarily time-dependent. We derive the spectral and Schmidt decompositions 
for vacuum solution. Using the decompositions we derive the analytical expressions for von Neumann and R\'{e}nyi entropies. Making use of Wigner distribution function defined in phase space, we derive the time-dependence of  position-momentum uncertainty relations. 
In order to show the dynamics of entanglement and uncertainty relation graphically we 
introduce two toy models and one realistic quenched model. While the dynamics can be conjectured by simple consideration in the toy models, the dynamics in the realistic quenched model is somewhat different from that in the toy models. In particular, the dynamics of entanglement exhibits similar pattern to dynamics of uncertainty parameter in the realistic quenched model. 
\end{abstract}

\maketitle

\section{Introduction}
Nickname of quantum entanglement\cite{schrodinger-35,text,horodecki09} is `spooky action at a distance' due to 
EPR paradox\cite{epr-35}. Although the debate related to  EPR paradox does seem to be far from complete conclusion, many theorists use the entanglement as
a physical resource to develop the various quantum information processing such as  quantum teleportation\cite{teleportation},
superdense coding\cite{superdense}, quantum cloning\cite{clon}, and quantum cryptography\cite{cryptography,cryptography2}. It is also quantum entanglement, which makes the quantum computer outperform the classical one\cite{qcreview,computer}. Furthermore, many experimentalists have tried to 
realize such quantum information processing in the laboratory for last decade. In particular, quantum cryptography seems to approaching to the 
commercial level\cite{white}.

In addition to quantum technology quantum entanglement is important notion in various branches of physics. The von Neumann\cite{woot-98} and 
R\'{e}nyi entropies\cite{renyi96}, which are frequently used to measure the bipartite entanglement, enable us to understand the Hawking-Bekenstein 
entropy\cite{bekenstein73,hawking76,hooft85,luca86,mark93,solo11} of black holes more deeply. They are also important to study on the 
quantum criticality\cite{eisert10,vidal03} and topological matters\cite{levin06,jiang12}.

Another important cornerstone in quantum mechanics is a uncertainty relation\cite{cohen}, which arises due to wave-particle dual property in the isolated systems. In this paper we examine the dynamics of the entanglement and uncertainty relation in coupled harmonic oscillator system, where the 
angular frequencies and coupling constant are arbitrarily time-dependent. The harmonic oscillator system is used in many branches of physics due to the fact 
that its mathematical simplicity provides a clear illustration of abstract ideas. For example, this system was used in Ref. \cite{han97} to discuss 
on the effect of the rest of universe\cite{feyman72}. It was shown that ignoring the rest of universe appears as an increase of uncertainty and entropy in the 
system in which we are interested. The analytical expression of von Neumann entropy was derived for a general real Gaussian density matrix in 
Ref. \cite{luca86} and it was generalized to massless scalar field in Ref. \cite{mark93}. Putting the scalar field system in the spherical box, the author 
in Ref. \cite{mark93} has shown that the total entropy of the system is proportional to surface area. This result gives some insight into a question why 
the Hawking-Bekenstein entropy is proportional to the area of the event horizon. Recently, the entanglement is computed in the coupled harmonic oscillator  system using a Schmidt decomposition\cite{maka17}. The von Neumann and R\'{e}nyi entropies are also explicitly computed in the similar system, called two site Bose-Hubbard model\cite{ghosh17}. The coupled harmonic oscillator system is also used in other branches such as 
molecular chemistry\cite{ikeda99,fillaux05} and biophysics\cite{bio1,bio2}.

This paper is organized as follows. In next section the diagonalization of Hamiltonian is discussed briefly. In Sec. III we derive the solutions for
 time-dependent Schr\"{o}dinger equation (TDSE) explicitly in the coupled harmonic oscillator system. In Sec. IV we derive the spectral and Schmidt decompositions
 for the vacuum solution. Using the decompositions we derive von Neumann and R\'{e}nyi entropies analytically if the oscillators are in the 
 ground states initially. In Sec. V we discuss on the dynamics of position-momentum uncertainty relation by making use of Wigner distribution function. 
 In Sec. VI we introduce two toy models and one 
 realistic quenched model, and derive the time-dependence of entanglement and uncertainty relation explicitly. It is shown that in the quenched model the pattern of uncertainty is similar to that of entanglement. In Sec. VII a brief conclusion is given. 
In appendix A the dynamics in the excited states is discussed briefly by assuming that the two oscillators are in ground and first-excited states initially.

\section{Diagonalization of Hamiltonian}
Let us consider the following Hamiltonian of coupled harmonic oscillator system
\begin{equation}
\label{hamil-1}
H = \frac{1}{2} (p_1^2 + p_2^2) + \frac{1}{2} \left( \omega_1^2 (t) x_1^2 + \omega_2^2 (t) x_2^2 \right) - J (t) x_1 x_2
\end{equation}
where $\{x_i, p_i \} \hspace{.1cm}  (i=1, 2)$ are the canonical coordinates and momenta, and  frequencies $\omega_j \hspace{.1cm} (j=1, 2)$ and coupling 
parameter $J$ are arbitrarily dependent on time. For simplicity, we assume that the oscillators have unit masses. Now, we define a rotation angle $\alpha$
as
\begin{eqnarray}
\label{define-1}
\left( \begin{array}{c} y_1  \\  y_2   \end{array}  \right) =    \left(    \begin{array}{cc}
                                                                                                         \cos \alpha  &  -\sin \alpha    \\
                                                                                                         \sin \alpha   &  \cos \alpha
                                                                                                                 \end{array}                  \right)
\left( \begin{array}{c} x_1  \\  x_2   \end{array}  \right).
\end{eqnarray}
If we choose  $\alpha$ as 
\begin{equation}
\label{angle}
\alpha = \frac{1}{2} \tan^{-1} \left( \frac{2 J}{\omega_1^2 - \omega_2^2} \right)
\end{equation}
with $-\pi / 4 \leq \alpha \leq \pi / 4$, the Hamiltonian is diagonalized as a form:
\begin{equation}
\label{hamil-2}
H = \frac{1}{2} \left( \tilde{p}_1^2 + \tilde{p}_2^2 \right) + \frac{1}{2} \left( \tilde{\omega}_1^2 (t) y_1^2 + \tilde{\omega}_2^2 (t) y_2^2 \right)
\end{equation}
where 
\begin{eqnarray}
\label{freq-1}
&&\tilde{\omega}_1^2 = \omega_1^2 + J \tan \alpha = \frac{1}{2} \left[ \left( \omega_1^2 + \omega_2^2 \right) + \epsilon (\omega_1^2 - \omega_2^2) \sqrt{(\omega_1^2 - \omega_2^2)^2 + 4 J^2} \right]    \\   \nonumber
&&\tilde{\omega}_2^2 = \omega_2^2 - J \tan \alpha = \frac{1}{2} \left[ \left( \omega_1^2 + \omega_2^2 \right) - \epsilon (\omega_1^2 - \omega_2^2) \sqrt{(\omega_1^2 - \omega_2^2)^2 + 4 J^2} \right]
\end{eqnarray}
with $\epsilon(x) = x / |x|$. Of course, $\tilde{p}_j = -i \partial / \partial y_j \hspace{.1cm} (j=1,2)$ are canonical momenta of $y_j$. In next section we will use the diagonalized Hamiltonian (\ref{hamil-2}) to solve 
the TDSE of the original Hamiltonian (\ref{hamil-1}).

\section{solutions of TDSE}
Consider a Hamiltonian of single harmonic oscillator with time-dependent frequency
\begin{equation}
\label{hamil-3}
H = \frac{p^2}{2} + \frac{1} {2} \omega^2 (t) x^2.
\end{equation}
The TDSE of this system was exactly solved in Ref. \cite{lewis68,lohe09}. The linearly independent solutions $\psi_n (x, t) \hspace{.1cm} (n=0, 1, \cdots)$ are expressed in a form
\begin{equation}
\label{TDSE-1}
\psi_n (x, t) = e^{-i E_n \tau(t)} e^{\frac{i}{2} \left( \frac{\dot{b}}{b} \right) x^2} \phi_n \left( \frac{x}{b} \right)
\end{equation}
where
\begin{eqnarray}
\label{TDSE-2}
&& E_n = \left( n + \frac{1}{2} \right) \omega(0)     \hspace{1.0cm}  \tau (t) = \int_0^t \frac{d s}{b^2 (s)}       \\   \nonumber
&&\phi_n (x) = \frac{1}{\sqrt{2^n n!}} \left( \frac{ \omega (0)} {\pi b^2} \right)^{1/4} H_n \left(\sqrt{\omega (0)} x \right) e^{-\frac{\omega (0)}{2} x^2 }.
\end{eqnarray}
In Eq. (\ref{TDSE-2}) $H_n (z)$ is $n^{th}$-order Hermite polynomial and $b(t)$ satisfies the Ermakov equation
\begin{equation}
\label{ermakov-1}
\ddot{b} + \omega^2 (t) b = \frac{\omega^2 (0)}{b^3}
\end{equation}
with $b(0) = 1$ and $\dot{b} (0) = 0$. Solution of the Ermakov equation was discussed in Ref. \cite{pinney50}. If $\omega(t)$ is time-independent, $b(t)$ is simply one. 
If $\omega (t)$ is instantly changed as
\begin{eqnarray}
\label{instant-1}
\omega (t) = \left\{                \begin{array}{cc}
                                               \omega_i  & \hspace{1.0cm}  t = 0   \\
                                               \omega_f  & \hspace{1.0cm}  t > 0,
                                               \end{array}            \right.
\end{eqnarray}
then $b(t)$ becomes
\begin{equation}
\label{scale-1}
b(t) = \sqrt{ \frac{\omega_f^2 - \omega_i^2}{2 \omega_f^2} \cos (2 \omega_f t) +  \frac{\omega_f^2 + \omega_i^2}{2 \omega_f^2}}.
\end{equation}
Recently. the solution (\ref{scale-1}) is extensively used to discuss the entanglement dynamics for the sudden quenched states of two site Bose-Hubbard model in Ref. \cite{ghosh17}.
Since TDSE is a linear differential equation, the general solution of TDSE is $\Psi (x, t) = \sum_{n=0}^{\infty} c_n \psi_n (x, t)$ with $\sum_{n=0}^{\infty} |c_n|^2 = 1$. The coefficient $c_n$ is 
determined by making use of the initial condition.

Using Eqs. (\ref{hamil-2}) and (\ref{TDSE-1}) the general solution for  TDSE of the  coupled harmonic oscillators is $\Psi (x_1, x_2 : t) = \sum_n \sum_m c_{n,m} \psi_{n,m} (x_1, x_2 : t)$, where
$\sum_n \sum_m |c_{n, m}|^2 = 1$ and 
\begin{eqnarray}
\label{solu-1}
&&\psi_{n,m} (x_1, x_2 : t) = \frac{1}{\sqrt{2^{n+m} n! m!}} \left( \frac{\omega'_1 \omega'_2} {\pi^2} \right)^{1/4}   \mbox{Exp} \Bigg[-i (E_n \tau_1 + E_m \tau_2)         \\     \nonumber
&&\hspace{2.0cm}                                                   - \frac{\omega'_1}{2}  (x_1 \cos \alpha - x_2 \sin \alpha )^2 - \frac{\omega'_2}{2} (x_1 \sin \alpha + x_2 \cos \alpha )^2         \\     \nonumber              
&&\hspace{2.0cm}+ \frac{i}{2} \left\{ \left( \frac{\dot{b_1}}{b_1} \right) (x_1 \cos \alpha - x_2 \sin \alpha )^2  + \left( \frac{\dot{b_2}}{b_2} \right) (x_1 \sin \alpha + x_2 \cos \alpha )^2 \right\} \Bigg]   \\   \nonumber
&&\hspace{2.0cm} \times H_n \left[\sqrt{\omega'_1} (x_1 \cos \alpha - x_2 \sin \alpha) \right]  H_n \left[\sqrt{\omega'_2} (x_1 \sin \alpha + x_2 \cos \alpha) \right].
\end{eqnarray}
In Eq. (\ref{solu-1}) $b_j \hspace{.1cm} (j=1,2)$ satisfy the  Ermakov equations  $\ddot{b_j} + \tilde{\omega}_j^2 (t) b_j = \frac{\tilde{\omega}_j^2 (0)}{b_j^3}$ respectively, and 
\begin{equation}
\label{boso-1}
\tau_j = \int_0^t \frac{d s}{b_j^2 (s)}   \hspace{1.0cm}   \omega'_j = \frac{\tilde{\omega}_j (0)}{ b_j^2}.
\end{equation}
The corresponding density matrix is defined as 
\begin{equation}
\label{density-1}
\rho (x_1, x_2: x'_1, x'_2 : t) = \Psi (x_1, x_2 : t) \Psi^* (x'_1, x'_2 : t).
\end{equation}
If two oscillators are $n^{th}-$ and $m^{th}-$states initially, the density matrix becomes
\begin{eqnarray}
\label{density-2}
&& \rho_{n,m} (x_1, x_2: x'_1, x'_2 : t) = \psi_{n,m} (x_1, x_2 : t) \psi^*_{n, m} (x'_1, x'_2 : t)     \\   \nonumber
&& = \frac{\sqrt{\omega'_1 \omega'_2}}{2^{n+m} n! m! \pi} H_n \left[\sqrt{\omega'_1} (x_1 \cos \alpha - x_2 \sin \alpha) \right] H_n \left[\sqrt{\omega'_1} (x'_1 \cos \alpha - x'_2 \sin \alpha) \right]    \\    \nonumber
&&\hspace{2.0cm} \times H_m \left[\sqrt{\omega'_2} (x_1 \sin \alpha + x_2 \cos \alpha) \right] H_m \left[\sqrt{\omega'_2} (x'_1 \sin \alpha + x'_2 \cos \alpha) \right]                                                        \\    \nonumber
&&\times \mbox{Exp} \left[-\frac{x_1^2}{2} (v_1 \cos^2 \alpha + v_2 \sin^2 \alpha ) - \frac{x_2^2}{2} (v_1 \sin^2 \alpha + v_2 \cos^2 \alpha) + x_1 x_2 \sin \alpha \cos \alpha (v_1 - v_2) \right]                    \\    \nonumber
&&\times \mbox{Exp} \left[-\frac{{x'_1}^{2}}{2} (v_1^* \cos^2 \alpha + v_2^* \sin^2 \alpha ) - \frac{{x'_2}^2}{2} (v_1^* \sin^2 \alpha + v_2^* \cos^2 \alpha) + x'_1 x'_2 \sin \alpha \cos \alpha (v_1^* - v_2^*) \right]
\end{eqnarray}
where $v_j = \omega'_j - i \frac{\dot{b_j}}{b_j}$. In next section we will discuss on the entanglement of the vacuum state $\rho_{0,0} (x_1, x_2: x'_1, x'_2 : t)$.

\section{Entanglement}
In order to explore the entanglement of the vacuum states we will derive the Schmidt decomposition of $\psi_{0,0} (x_1, x_2: t)$ and the spectral decomposition of the reduced density matrix explicitly.
The reduced density matrix of the first oscillator is given by 
\begin{equation}
\label{density-3}
\rho_{(0,0)}^A (x_1, x'_1: t) \equiv \int d x_2 \rho_{0, 0} (x_1, x_2 : x'_1, x_2: t).
\end{equation}
The explicit expression of the reduced density matrix is 
\begin{equation}
\label{density-4}
\rho_{(0,0)}^A (x_1, x'_1: t) = \sqrt{\frac{2 a_1}{\pi}} \mbox{Exp} \left[ - \left\{ (a_1 + a_3) - i a_2 \right\} x_1^2 -  \left\{ (a_1 + a_3) + i a_2 \right\} {x'_1}^2 + 2 a_3 x_1 x'_1 \right]
\end{equation}
where 
\begin{equation}
\label{boso-2}
a_1 = \frac{\omega'_1 \omega'_2}{2 D} \hspace{.5cm}
a_2 = \frac{\omega'_1 \frac{\dot{b}_2}{b_2} \sin^2 \alpha + \omega'_2 \frac{\dot{b}_1}{b_1} \cos^2 \alpha}{2 D}    \hspace{.5cm}
a_3 = \frac{\sin^2 \alpha \cos^2 \alpha \left[ (\omega'_1 - \omega'_2)^2 + \left( \frac{\dot{b}_1}{b_1} -  \frac{\dot{b}_2}{b_2} \right)^2 \right]}{4 D}
\end{equation}
with $D = \omega'_1 \sin^2 \alpha + \omega'_2 \cos^2 \alpha$. One can show easily 
\begin{eqnarray}
\label{trace-1}
&&\mbox{Tr} \left[\rho_{(0,0)}^A \right] \equiv \int dx \rho_{(0,0)}^A (x, x: t) = 1        \\     \nonumber
&&\mbox{Tr} \left[\left(\rho_{(0,0)}^A \right)^2\right] \equiv \int dx dx' \rho_{(0,0)}^A (x, x': t) \rho_{(0,0)}^A (x', x: t) = \sqrt{\frac{a_1}{a_1 + 2 a_3}} = \sqrt{\frac{\omega'_1 \omega'_2}{\bar{\eta}}}
\end{eqnarray}
where $\bar{\eta} = D \tilde{D} + \sin^2 \alpha \cos^2 \alpha  \left( \frac{\dot{b}_1}{b_1} -  \frac{\dot{b}_2}{b_2} \right)^2$ with $\tilde{D} = \omega'_1 \cos^2 \alpha + \omega'_2 \sin^2 \alpha$.
First equation of Eq. (\ref{trace-1}) guarantees the probability conservation and second one denotes the mixedness of  $ \rho_{(0,0)}^A (x, x': t)$. If it is  one, this means that $\rho_{(0,0)}^A (x, x': t)$ is pure state. If, on the contrary,
it is zero, this means that $\rho_{(0,0)}^A (x, x': t)$ is completely mixed state. If $\tilde{\omega}_j$ are independent of time, $\omega'_j = \tilde{\omega}_j$ and $\mbox{Tr} \left[\left(\rho_{(0,0)}^A \right)^2\right]$ becomes $\sqrt{\omega'_1 \omega'_2 / (D \tilde{D})}$. Thus, 
if $\alpha = 0$,  $ \rho_{(0,0)}^A (x, x': t)$ becomes pure state. The most strong mixed states occur at $\alpha = \pm \pi / 4$, In this case 
 $\mbox{Tr} \left[\left(\rho_{(0,0)}^A \right)^2\right]$ becomes $2 \sqrt{\omega'_1 \omega'_2} / (\omega'_1 + \omega'_2)$.

In order to derive the spectral decomposition of $\rho_{(0,0)}^A (x, x': t)$ we should solve the following eigenvalue equation;
\begin{equation}
\label{eigen-1}
\int_{-\infty}^{\infty} dx' \rho_{(0,0)}^A (x, x': t) f_n (x', t) = p_n (t) f_n (x', t).
\end{equation}
Similar problem was discussed in Ref. \cite{ghosh17,luca86,mark93}. It is not difficult to show that  the eigenvalue and normalized  
eigenfunction in this case are
\begin{equation}
\label{eigen-2}
f_n (x, t) = \frac{1}{\sqrt{2^n n!}} \left(\frac{\epsilon}{\pi}\right)^{1/4} H_n (\sqrt{\epsilon} x) \mbox{Exp} \left[-\frac{\epsilon}{2} x^2 + i a_2 x^2 \right]     \hspace{.5cm}
 p_n(t) =(1 - \xi) \xi^n
\end{equation}
where 
\begin{equation}
\label{eigen-3}
\epsilon = 2 \sqrt{(a_1 + a_3)^2 - a_3^2}    \hspace{1.0cm}  \xi = \frac{a_3}{(a_1 + a_3) + \frac{\epsilon}{2}}.
\end{equation}
Thus, the spectral decomposition of $ \rho_{(0,0)}^A (x, x': t)$ can be written as 
\begin{equation}
\label{spectral-1}
 \rho_{(0,0)}^A (x, x': t) = \sum_{n=0}^{\infty} p_n (t) f_n (x, t) f_n^* (x', t).
 \end{equation}
 This can be proved explicitly by making use of a mathematical formula
 $$\sum_{n=0}^{\infty} \frac{t^n}{n!} H_n (x) H_n (y) = (1 - 4 t^2)^{-1/2} \mbox{Exp}\left[\frac{4 t x y - 4 t^2 (x^2 + y^2)}{1 - 4 t^2} \right].$$
 Thus R\'{e}nyi and von Neumann entropies are given by 
 \begin{eqnarray}
 \label{entropy-1}
 && S_{n} \equiv \frac{1}{1 - n} \ln \mbox{Tr} \left(\left(  \rho_{(0,0)}^A (x, x': t) \right)^n \right) = \frac{1}{1 - n} \ln \frac{(1 - \xi)^n}{1 - \xi^n}                            \\    \nonumber
 && S_{von} \equiv \lim_{n \rightarrow 1} S_n = - \ln (1 - \xi) - \frac{\xi}{1 - \xi} \ln \xi
 \end{eqnarray}
 where $n$ is any positive integer. It is worthwhile noting that when $\alpha = 0$, $\xi$ becomes zero which results in vanishing  R\'{e}nyi and von Neumann entropies. It is 
 obvious because $\alpha = 0$ corresponds to $J = 0$ and,  two oscillators are completely decoupled. 
 
 In order to derive the Schmidt decomposition of $\psi_{0,0} (x_1, x_2: t)$ we should solve the eigenvalue equation of other party.
 The reduced density matrix of other party is given by 
 \begin{eqnarray}
 \label{density-B-1}
&& \rho_{(0,0)}^B (x_2, x'_2: t) \equiv \int d x_1 \rho_{0, 0} (x_1, x_2 : x_1, x'_2: t)                  \\     \nonumber 
&&\hspace{1.0cm}= \sqrt{\frac{2 \tilde{a}_1}{\pi}} \mbox{Exp} \left[ - \left\{ (\tilde{a}_1 + \tilde{a}_3) - i \tilde{a}_2 \right\} x_2^2 -  \left\{ (\tilde{a}_1 + \tilde{a}_3) + i \tilde{a}_2 \right\} {x'_2}^2 + 2 \tilde{a}_3 x_2 x'_2 \right]
\end{eqnarray}
where
\begin{equation}
\label{bosoB-2}
\tilde{a}_1 = \frac{\omega'_1 \omega'_2}{2 \tilde{D}} \hspace{.5cm}
\tilde{a}_2 = \frac{\omega'_1 \frac{\dot{b}_2}{b_2} \cos^2 \alpha + \omega'_2 \frac{\dot{b}_1}{b_1} \sin^2 \alpha}{2 \tilde{D}}    \hspace{.5cm}
\tilde{a}_3 = \frac{\sin^2 \alpha \cos^2 \alpha \left[ (\omega'_1 - \omega'_2)^2 + \left( \frac{\dot{b}_1}{b_1} -  \frac{\dot{b}_2}{b_2} \right)^2 \right]}{4 \tilde{D}}.
\end{equation}
The eigenvalues of $\rho_{(0,0)}^B$ are exactly the same with those of $\rho_{(0,0)}^A$ and the eigenfunction becomes
\begin{equation}
\label{eigen-B-1}
\tilde{f}_n (x, t) = \frac{1}{\sqrt{2^n n!}} \left(\frac{\tilde{\epsilon}}{\pi}\right)^{1/4} H_n (\sqrt{\tilde{\epsilon}} x) \mbox{Exp} \left[-\frac{\tilde{\epsilon}}{2} x^2 + i \tilde{a}_2 x^2 \right]
\end{equation}
where $\tilde{\epsilon} = 2 \sqrt{(\tilde{a}_1 + \tilde{a}_3)^2 - \tilde{a}_3^2}$.
Then one can find the Schmidt decomposition of $\psi_{0,0} (x_1, x_2: t)$, which is 
\begin{equation}
\label{schmidt-1}
\psi_{0,0} (x_1, x_2: t) = \sum_n \sqrt{p_n} \left[ f_n (x_1, t) e^{-i n \theta / 2} e^{-i (E_0 \tau_1 - \varphi / 4)} \right] \left[ \tilde{f}_n (x_2, t) e^{-i n \theta / 2} e^{-i (E_0 \tau_2 - \varphi / 4)} \right] 
\end{equation}
where
\begin{equation}
\label{schmidt-2}
\theta = \tan^{-1} \left( \frac{Z_2}{Z_1} \right)   \hspace{1.0cm}
\varphi = \tan^{-1} \left( \frac{(\kappa - 1) Z_1 Z_2}{Z_1^2 + \kappa Z_2^2} \right).
\end{equation}
In Eq. (\ref{schmidt-2}) $Z_1$, $Z_2$, and $\kappa$ are given by 
\begin{eqnarray}
\label{schmidt-3}
&&\kappa = \left[ 1 + \frac{\sin^2 \alpha \cos^2 \alpha}{\omega'_1 \omega'_2} \left\{ (\omega'_1 - \omega'_2)^2 + \left(\frac{\dot{b}_1}{b_1} - \frac{\dot{b}_2}{b_2} \right)^2 \right\} \right]^{1/2}    \\   \nonumber
&& \hspace{2.0cm}Z_1 = \omega'_1 - \omega'_2            \hspace{1.0cm}  Z_2 = \frac{1}{\kappa} \left(\frac{\dot{b}_1}{b_1} - \frac{\dot{b}_2}{b_2} \right).
\end{eqnarray}
If $\tilde{\omega}_j$ are independent of time, $Z_2$ becomes zero, which results in $\theta = \varphi = 0$. 
From the Schmidt decomposition one can construct other bipartite entanglement measures such as St\"{u}ckelberg entropy. Furthermore, 
Schmidt basis is very useful to discuss on the entanglement in quantum optics, atom-field interaction, and electron-electron
correlation\cite{grobe94,ekert95}. In this paper, however, we consider only R\'{e}nyi  and von Neumann entropies as bipartite entanglement measures.

\section{Uncertainty Relations}
In order to discuss on the time-dependence of the uncertainty it is convenient to compute the Wigner distribution function defined 
\begin{equation}
\label{wigner-1}
W (x_1, x_2: p_1, p_2: t) \equiv \frac{1}{\pi^2} \int dy_1 dy_2 e^{-2 i (p_1 y_1 + p_2 y_2)} \Psi^* (x_1 + y_1, x_2 + y_2: t) \Psi (x_1 - y_1, x_2 - y_2: t).
\end{equation}
Many interesting properties of the Wigner function are discussed in Ref.\cite{feyman72,noz91}. In particular, it is convenient to introduce the Wigner 
distribution function in the density matrix formalism when we want to study the uncertainty relations in detail.

If we choose the wave function as $\psi_{0,0} (x_1, x_2: t)$, the Wigner function becomes
\begin{eqnarray}
\label{wigner-2}
&&W_{(0,0)} (x_1, x_2: p_1, p_2: t)  = \frac{1}{\pi^2} \mbox{Exp} \bigg[ -A_1 x_1^2 - A_2 x_2^2 - B_1 p_1^2 - B_2 p_2^2                 \\    \nonumber
&&\hspace{1.0cm}+ 2 A_3 x_1 x_2 + 2 B_3 p_1 p_2 + 2 F (x_1 p_2 + x_2 p_1) + 2 D_{11} x_1 p_1 + 2 D_{22} x_2 p_2 \bigg]         \\    \nonumber
  \end{eqnarray}
where
\begin{eqnarray}
\label{wigner-3}
&&A_1 = \frac{1}{\omega'_1 \omega'_2} \left[ \omega'_1 \omega'_2 \tilde{D} + \omega'_2 \left(\frac{\dot{b}_1}{b_1}\right)^2 \cos^2 \alpha + \omega'_1 \left( \frac{\dot{b}_2}{b_2} \right)^2 \sin^2 \alpha \right]    \\   \nonumber
&&A_2 = \frac{1}{\omega'_1 \omega'_2} \left[ \omega'_1 \omega'_2 D + \omega'_2 \left(\frac{\dot{b}_1}{b_1}\right)^2 \sin^2 \alpha + \omega'_1 \left( \frac{\dot{b}_2}{b_2} \right)^2 \cos^2 \alpha \right]               \\    \nonumber
&&A_3 = \frac{\sin \alpha \cos \alpha}{\omega'_1 \omega'_2} \left[ \omega'_1 \omega'_2 (\omega'_1 - \omega'_2)  + \omega'_2 \left(\frac{\dot{b}_1}{b_1}\right)^2 - \omega'_1 \left( \frac{\dot{b}_2}{b_2} \right)^2\right]    \\   \nonumber
&& B_1 = \frac{D}{\omega'_1 \omega'_2}     \hspace{1.0cm}  B_2 = \frac{\tilde{D}}{\omega'_1 \omega'_2} \hspace{1.0cm} B_3 = - \frac{\sin \alpha \cos \alpha}{\omega'_1 \omega'_2} (\omega'_1 - \omega'_2)                      \\   \nonumber
&& F = \frac{\sin \alpha \cos \alpha}{\omega'_1 \omega'_2} \left(\omega'_2 \frac{\dot{b}_1}{b_1} - \omega'_1 \frac{\dot{b}_2}{b_2} \right) \hspace{1.0cm} 
D_{11} = - \frac{1}{\omega'_1 \omega'_2}  \left(\omega'_2 \frac{\dot{b}_1}{b_1} \cos^2 \alpha + \omega'_1 \frac{\dot{b}_2}{b_2} \sin^2 \alpha \right)                                                                                                                    \\   \nonumber
&&\hspace{2.0cm} D_{22} = - \frac{1}{\omega'_1 \omega'_2}  \left(\omega'_2 \frac{\dot{b}_1}{b_1} \sin^2 \alpha + \omega'_1 \frac{\dot{b}_2}{b_2} \cos^2 \alpha \right).
\end{eqnarray}
The Wigner function $W_{(0,0)} (x_1, p_1: t)$ is defined from $W_{(0,0)} (x_1, x_2: p_1, p_2: t)$ as 
\begin{equation}
\label{wigner-4}
W_{(0,0)} (x_1, p_1: t) = \int dx_2 dp_2 W_{(0,0)} (x_1, x_2: p_1, p_2: t).
\end{equation}
Using Eq. (\ref{wigner-2}) one can show 
\begin{equation}
\label{wigner-5}
W_{(0,0)} (x_1, p_1: t) = \frac{1}{\pi} \sqrt{\frac{\omega'_1 \omega'_2}{\bar{\eta}}} e^{-\alpha_1 x_1^2 - \alpha_2 p_1^2 + 2 \alpha_3 x_1 p_1}
\end{equation}
where
\begin{eqnarray}
\label{wigner-6}
&&\alpha_1 = \frac{1}{\bar{\eta}} \left[ \tilde{D} \omega'_1 \omega'_2 + \omega'_2 \left(\frac{\dot{b}_1}{b_1}\right)^2 \cos^2 \alpha + \omega'_1 \left( \frac{\dot{b}_2}{b_2} \right)^2 \sin^2 \alpha \right]                \\    \nonumber
&& \alpha_2 = \frac{D}{\bar{\eta}}    \hspace{1.0cm}
\alpha_3 = -  \frac{1}{\bar{\eta}} \left( \omega'_2 \frac{\dot{b}_1}{b_1} \cos^2 \alpha + \omega'_1  \frac{\dot{b}_2}{b_2} \sin^2 \alpha \right).
\end{eqnarray}
It is worthwhile noting that $\alpha_j \hspace{.1cm} (j=1,2,3)$ satisfy
$$ \alpha_1 \alpha_2 - \alpha_3^2 = \frac{\omega'_1 \omega'_2}{\bar{\eta}}.$$
One can show straightforwardly 
\begin{eqnarray}
\label{wigner-7}
&&\int dx_1 dp_1 W_{(0,0)} (x_1, p_1: t) = 1 = \mbox{Tr} \left[\rho_{(0,0)}^A \right]      \\    \nonumber
&&2 \pi \int dx_1 dp_1 W_{(0,0)}^2 (x_1, p_1: t) = \sqrt{\frac{\omega'_1 \omega'_2}{\bar{\eta}}} = \mbox{Tr} \left[\left(\rho_{(0,0)}^A \right)^2\right].
\end{eqnarray}

In terms of the Wigner distribution function the average of a quantity ${\cal O} (x_1, p_1)$ is defined as 
\begin{equation}
\label{average-1}
<{\cal O}> (x_1, p_1) \equiv \int {\cal O} (x_1, p_1) W_{(0,0)} (x_1, p_1) dx_1 dp_1.
\end{equation}
Then it is straightforward to show that $<x_1> = <p_1> = 0$ and 
\begin{equation}
\label{average-2}
<x_1^2> = \frac{D}{2 \omega'_1 \omega'_2}  
\hspace{1.0cm} <p_1^2> = \frac{1}{2} \left[ \tilde{D} + \frac{1}{\omega'_1} \left(\frac{\dot{b}_1}{b_1}\right)^2 \cos^2 \alpha + \frac{1}{\omega'_2} \left( \frac{\dot{b}_2}{b_2} \right)^2 \sin^2 \alpha \right].
\end{equation}
Thus, the position-momentum uncertainty for the vacuum state becomes
\begin{equation}
\label{uncertainty-1}
\left(\Delta x_1 \Delta p_1 \right)^2 = \frac{1}{4} \Omega (t)
\end{equation}
where 
\begin{equation}
\label{uncertainty-2}
\Omega (t) = \left( \frac{1}{\omega'_1} \cos^2 \alpha + \frac{1}{\omega'_2} \sin^2 \alpha \right) \left[ \left\{ \omega'_1 + \frac{1}{\omega'_1} \left(\frac{\dot{b}_1}{b_1}\right)^2 \right\} \cos^2 \alpha
+ \left\{ \omega'_2 + \frac{1}{\omega'_2} \left( \frac{\dot{b}_2}{b_2} \right)^2 \right\} \sin^2 \alpha \right].
\end{equation}
Using Eq. (\ref{density-B-1}) one can compute also the uncertainty between $x_2$ and $p_2$. In this case the uncertainty becomes $(\Delta x_2 \Delta p_2)^2 = \tilde{\Omega} (t) / 4$, where 
\begin{equation}
\label{uncertainty-3}
\tilde{\Omega} (t) = \left( \frac{1}{\omega'_1} \sin^2 \alpha + \frac{1}{\omega'_2} \cos^2 \alpha \right) \left[ \left\{ \omega'_1 + \frac{1}{\omega'_1} \left(\frac{\dot{b}_1}{b_1}\right)^2 \right\} \sin^2 \alpha
+ \left\{ \omega'_2 + \frac{1}{\omega'_2} \left( \frac{\dot{b}_2}{b_2} \right)^2 \right\} \cos^2 \alpha \right].
\end{equation}
If $\tilde{\omega}_j$ are time-independent, both $\Omega$ and $\tilde{\Omega}$ reduces to $1 + \frac{ (\omega'_1 - \omega'_2)^2}{\omega'_1 \omega'_2}  \sin^2 \alpha \cos^2 \alpha $. Thus minimum uncertainty occurs at $\alpha = 0$
while maximum uncertainty occurs at $\alpha = \pm \pi / 4$. 

\section{Dynamics of Entanglement and Uncertainty}

%%%%%%%%%%%%%%%%%%%%%%%%%%%%%%%%%%%%%%%%%%%%%%%%%%%%%%%%%
\begin{figure}[ht!]
\begin{center}
\includegraphics[height=5.0cm]{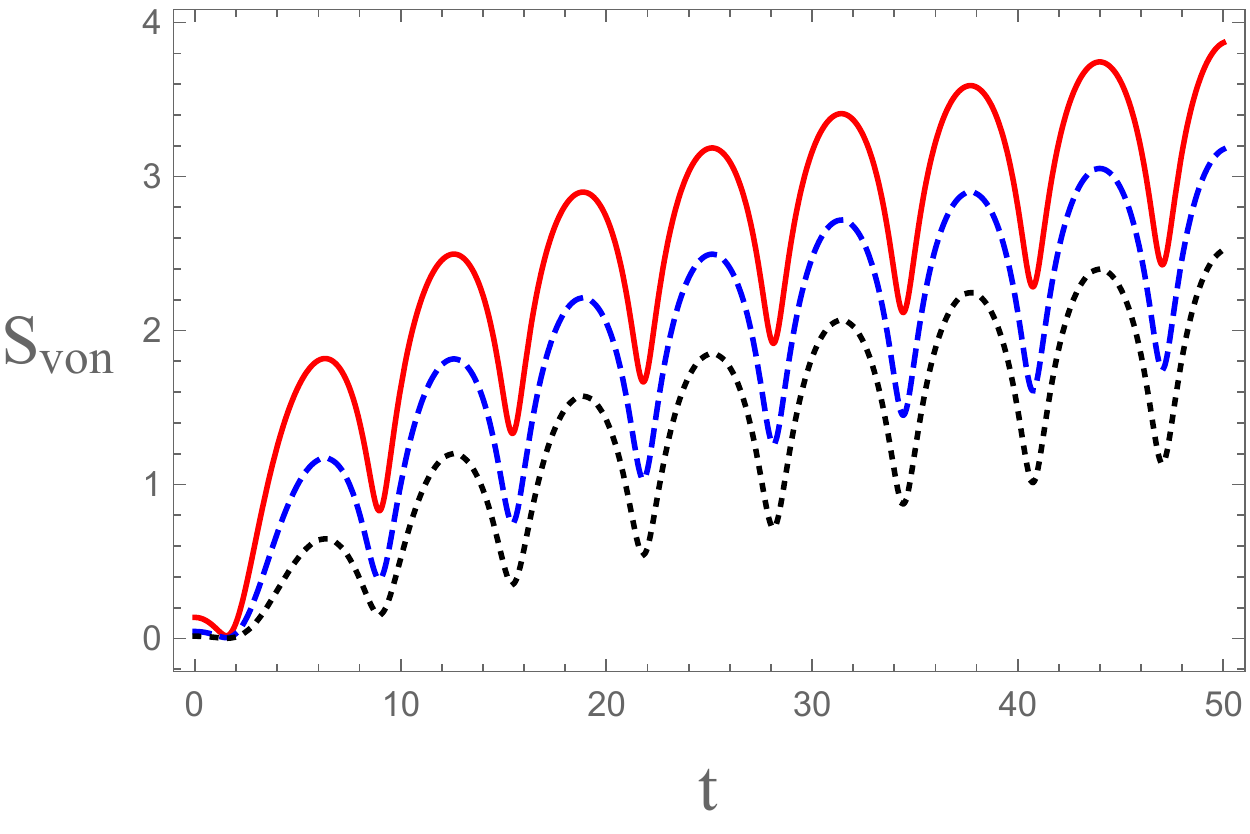} \hspace{0.2cm}
\includegraphics[height=5.0cm]{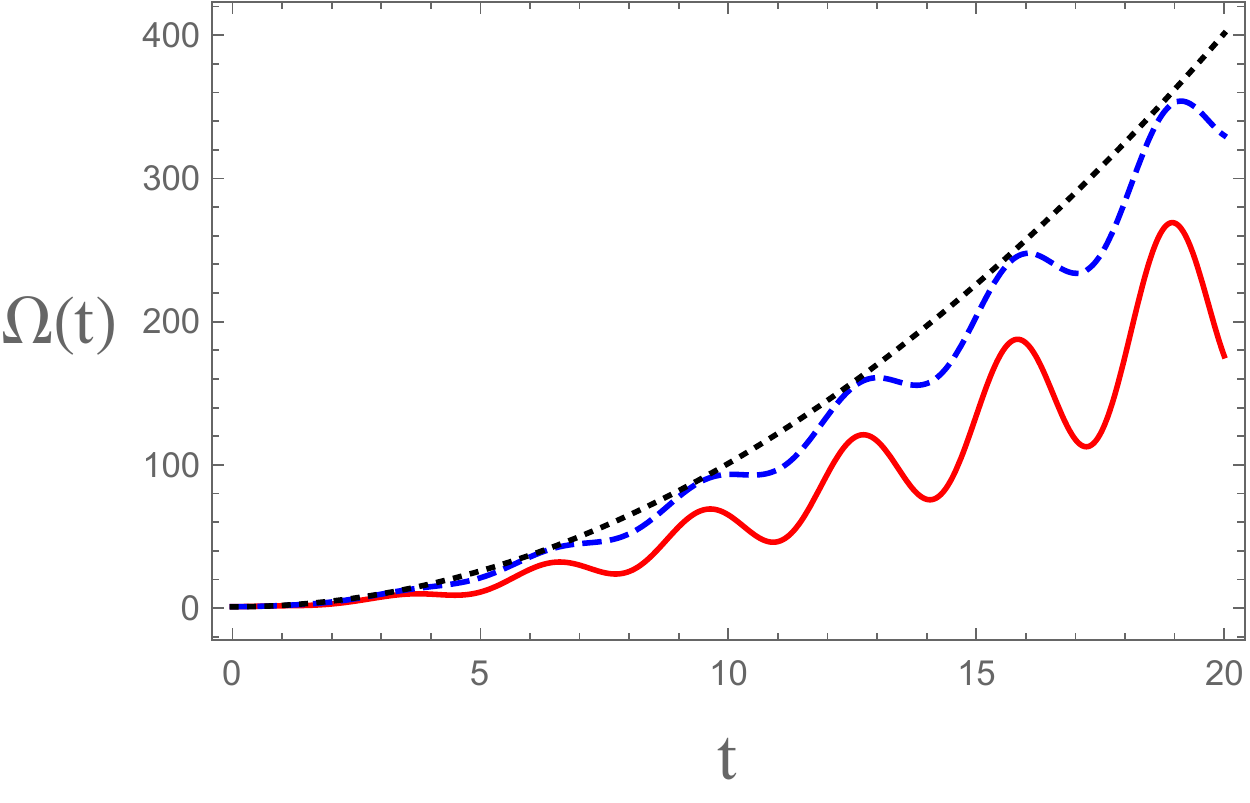}

\caption[fig1]{(Color online) The time-dependence of von Neumann entropy (Fig. 1(a)) and uncertainty $\Omega (t)$ (Fig. 1(b))when $\omega_{1, i} = 1$, $\omega_{1, f} = 0$, $\omega_{2, i} = 2$, and $\omega_{2, f} = 0.5$ for various $\alpha$ in the first model. (a) Dynamics of von Neumann entropy for $\alpha = \pi / 4$ (red solid line), 
$\alpha = \pi / 12$ (blue dashed line), and $\alpha = \pi / 24$ (black dotted line). (b)  Dynamics of uncertainty for $\alpha = \pi / 4$ (red solid line), 
$\alpha = \pi / 8$ (blue dashed line), and $\alpha = 0$ (black dotted line). }
\end{center}
\end{figure}
%%%%%%%%%%%%%%%%%%%%%%%%%%%%%%%%%%%%%%%%%%%%%%%%%%%%%%%%%%%

In this section we examine the dynamics of entanglement and uncertainty in three models. First two models are toy models, which are 
introduced to examine the effect of time-dependence of the angular frequencies and rotation angle $\alpha$ in the dynamics. As third model we 
introduce more realistic quenched model, where the dynamics can be solved analytically. Although we can consider more general case by solving the Ermakov
equation (\ref{ermakov-1}) numerically, this fully general model is not explored in this paper because we would like to confine ourselves to analytic
cases.

The first model we consider is a simple case that one of the angular frequencies $\tilde{\omega}_j$ is zero at late time. We choose 
\begin{eqnarray}
\label{model1-1}
\tilde{\omega}_1 (t) = \left\{                \begin{array}{cc}
                                               \omega_{1, i}  & \hspace{0.25cm} t = 0   \\
                                                \omega_{1, f} = 0   & \hspace{0.5cm} t > 0
                                               \end{array}            \right.         \hspace{1.0cm}
\tilde{\omega}_2 (t) = \left\{                \begin{array}{cc}
                                               \omega_{2, i}  & \hspace{0.25cm} t = 0   \\
                                                \omega_{2, f}   & \hspace{0.25cm}  t > 0.
                                               \end{array}            \right.
\end{eqnarray}
From Eq. (\ref{freq-1}) this is achieved by $\omega_1 \omega_2 = \pm J$ with $\omega_2^2 > \omega_1^2$ at $t > 0$.
In this case $b_1 (t)$ and $b_2(t)$ become
\begin{equation}
\label{model1-2}
b_1 (t) = \sqrt{1 +  \omega_{1, i}^2 t^2}  \hspace{1.0cm}
b_2 (t) = \sqrt{ \left(\frac{\omega_{2, f}^2 - \omega_{2, i}^2}{2 \omega_{2, f}^2}\right) \cos (2 \omega_{2, f} t) + \left( \frac{\omega_{2, f}^2 + \omega_{2, i}^2}{2 \omega_{2, f}^2}\right)}.
\end{equation}
The time-dependence of the von Neumann entropy for $\alpha = \pi / 4$ (red solid line),  $\alpha = \pi / 12$ (blue dashed line), and 
$\alpha = \pi / 24$ (black dotted line) is plotted  in Fig. 1(a) 
when $\omega_{1, i} = 1$, $\omega_{1, f} = 0$, $\omega_{2, i} = 2$, and $\omega_{2, f} = 0.5$. It exhibits an increasing behavior with oscillation. This oscillation is mainly due to 
$b_2 (t)$. The figure shows that the coupled  harmonic oscillator is more entangled with increasing $|\alpha|$. This can be expected from the fact that the 
oscillators become separable when $\alpha = 0$. The time-dependence of uncertainty $\Omega (t) = (2 \Delta x_1 \Delta p_1)^2$ is plotted in Fig. 1(b) for $\alpha = \pi / 4$ (red solid line), $\alpha = \pi / 8$ (blue dashed line) and $\alpha = 0$ (black dotted line). The uncertainty is maximized at the separable oscillator system and is minimized at $|\alpha| = \pi /4$ at most domain of time. However, 
this order is reversed at the small $t$ region (for our case $0 < t < 0.773$). In this region the uncertainty is maximized at $\alpha = \pi / 4$ and is minimized 
at $\alpha = 0$. The oscillatory behavior is also due to $b_2(t)$. 

%%%%%%%%%%%%%%%%%%%%%%%%%%%%%%%%%%%%%%%%%%%%%%%%%%%%%%%%%
\begin{figure}[ht!]
\begin{center}
\includegraphics[height=5.0cm]{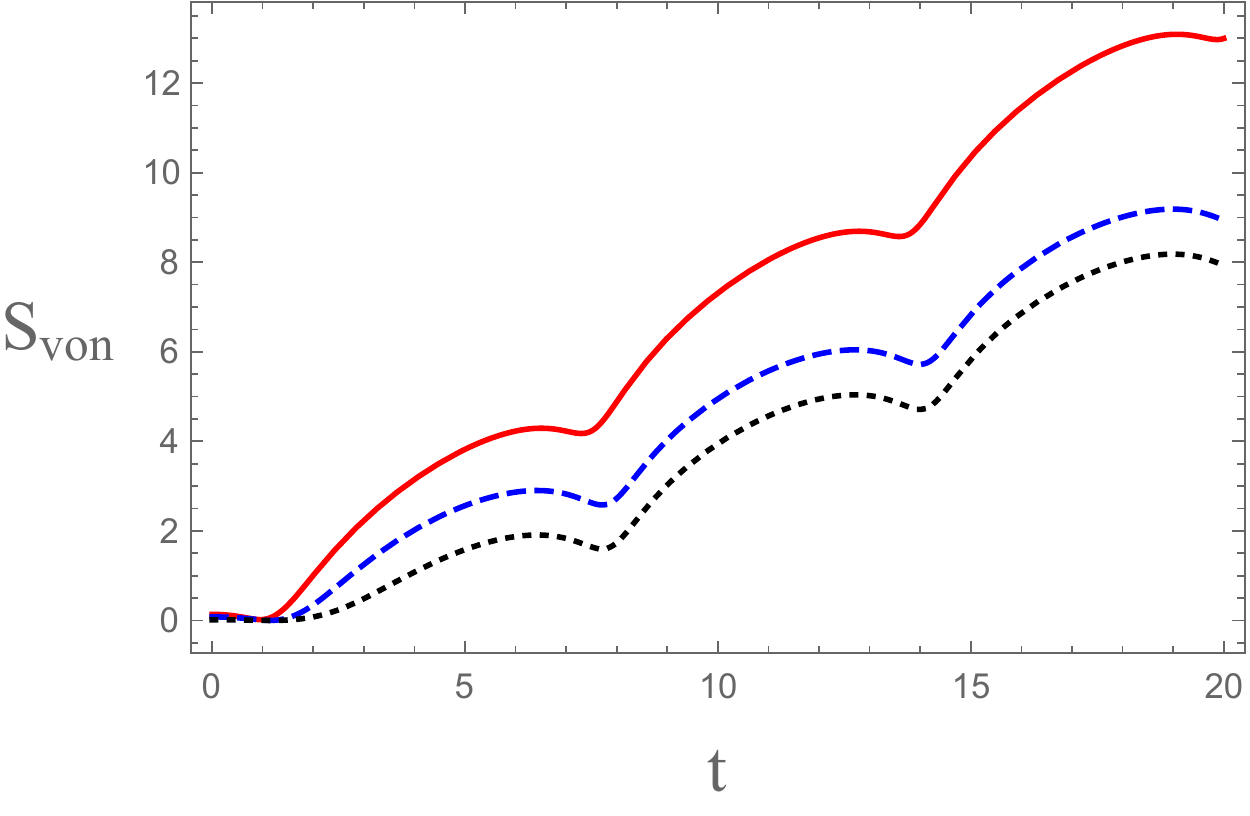} \hspace{0.2cm}
\includegraphics[height=5.0cm]{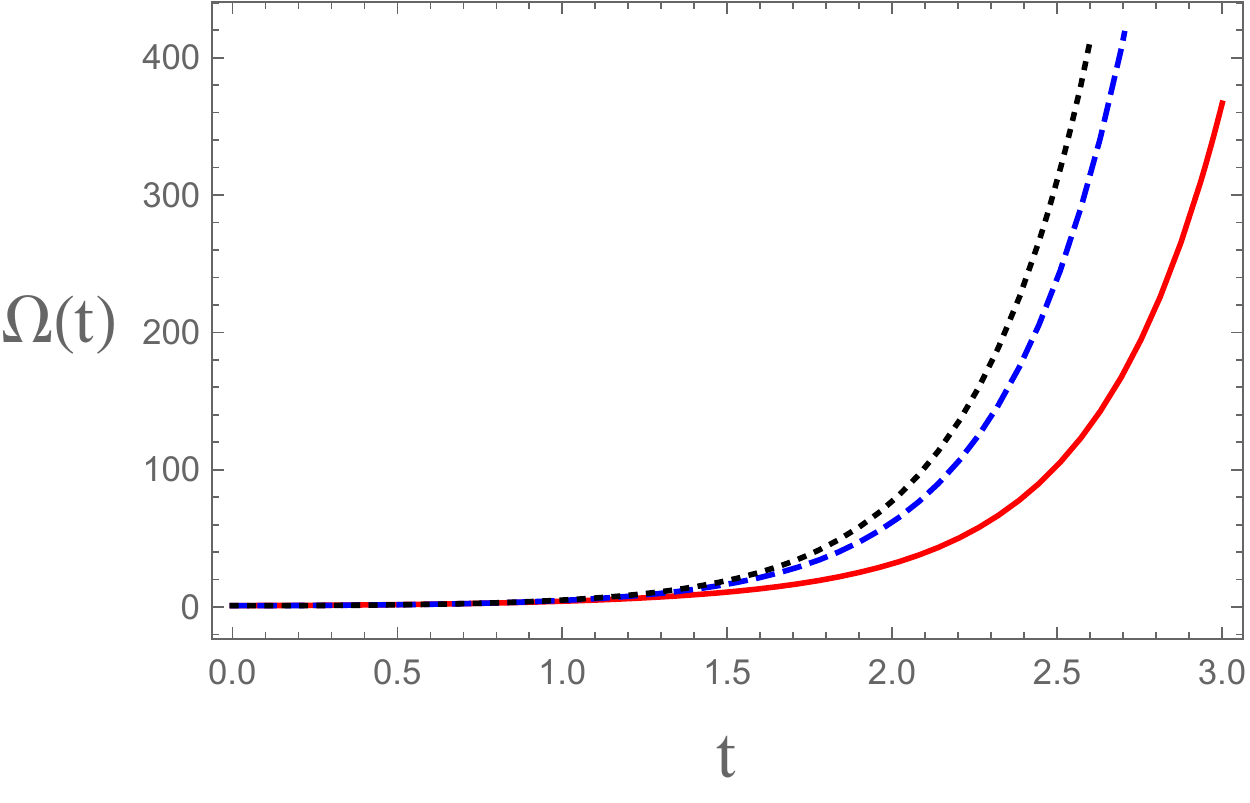}

\caption[fig2]{(Color online) The time-dependence of von Neumann entropy (Fig. 2(a)) and uncertainty $\Omega (t)$ (Fig. 2(b))when $\omega_{1, i} = 1$, $\omega_{1, f} = 0.7 $, $\omega_{2, i} = 2$, and $\omega_{2, f} = 0.5$ for various $\alpha$ in the second model. (a) Dynamics of von Neumann entropy for $\alpha = \pi / 4$ (red solid line), 
$\alpha = \pi / 8$ (blue dashed line), and $\alpha = \pi / 24$ (black dotted line). (b)  Dynamics of uncertainty for $\alpha = \pi / 4$ (red solid line), 
$\alpha = \pi / 8$ (blue dashed line), and $\alpha = 0$ (black dotted line). }
\end{center}
\end{figure}
%%%%%%%%%%%%%%%%%%%%%%%%%%%%%%%%%%%%%%%%%%%%%%%%%%%%%%%%%%%

Second simple model we consider is a case that one of the angular frequencies $\tilde{\omega}_j$ is imaginary at late time.  We choose 
\begin{eqnarray}
\label{model2-1}
\tilde{\omega}_1 (t) = \left\{                \begin{array}{cc}
                                               \omega_{1, i}  & \hspace{0.25cm} t = 0   \\
                                                i \omega_{1, f}   & \hspace{0.5cm} t > 0
                                               \end{array}            \right.         \hspace{1.0cm}
\tilde{\omega}_2 (t) = \left\{                \begin{array}{cc}
                                               \omega_{2, i}  & \hspace{0.25cm} t = 0   \\
                                                \omega_{2, f}   & \hspace{0.25cm}  t > 0.
                                               \end{array}            \right.
\end{eqnarray}
From Eq. (\ref{freq-1}) this is achieved by $J^2 > \omega_1^2 \omega_2^2$ with $\omega_2^2 > \omega_1^2$ at $t > 0$.
In this case $b_2(t)$ is not changed and $b_1(t)$ becomes 
\begin{equation}
\label{model2-2}
b_1 (t) = \sqrt{ \left(\frac{\omega_{1, f}^2 + \omega_{1, i}^2}{2 \omega_{1, f}^2}\right) \cosh (2 \omega_{1, f} t) + \left( \frac{\omega_{1, f}^2 - \omega_{1, i}^2}{2 \omega_{1, f}^2}\right)}.
\end{equation}
The time-dependence of the von Neumann entropy for $\alpha = \pi / 4$ (red solid line),  $\alpha = \pi / 8$ (blue dashed line), and 
$\alpha = \pi / 24$ (black dotted line) is plotted  in Fig. 2(a) 
when $\omega_{1, i} = 1$, $\omega_{1, f} = 0.7 $, $\omega_{2, i} = 2$, and $\omega_{2, f} = 0.5$. Like a previous case 
it exhibits an increasing behavior with oscillation. The difference is the fact that the von Neumann entropy in the present case is rapidly increasing in time compared to the previous case. This seems to be mainly due to exponential behavior of $b_1(t)$ in time. 
The time-dependence of uncertainty $\Omega (t) = (2 \Delta x_1 \Delta p_1)^2$ is plotted in Fig. 2(b) for $\alpha = \pi / 4$ (red solid line), $\alpha = \pi / 8$ (blue dashed line) and $\alpha = 0$ (black dotted line). Although whole behavior is similar to the previous case, the oscillatory behavior disappears in this case. This is due to the rapid increasing behavior of $\Omega (t)$, thus the amplitude of oscillation is negligible. In this case also the order of uncertainty is 
reversed at the small $t$ region (for this case $0 \leq t \leq 0.713$).

%%%%%%%%%%%%%%%%%%%%%%%%%%%%%%%%%%%%%%%%%%%%%%%%%%%%%%%%%
\begin{figure}[ht!]
\begin{center}
\includegraphics[height=5.0cm]{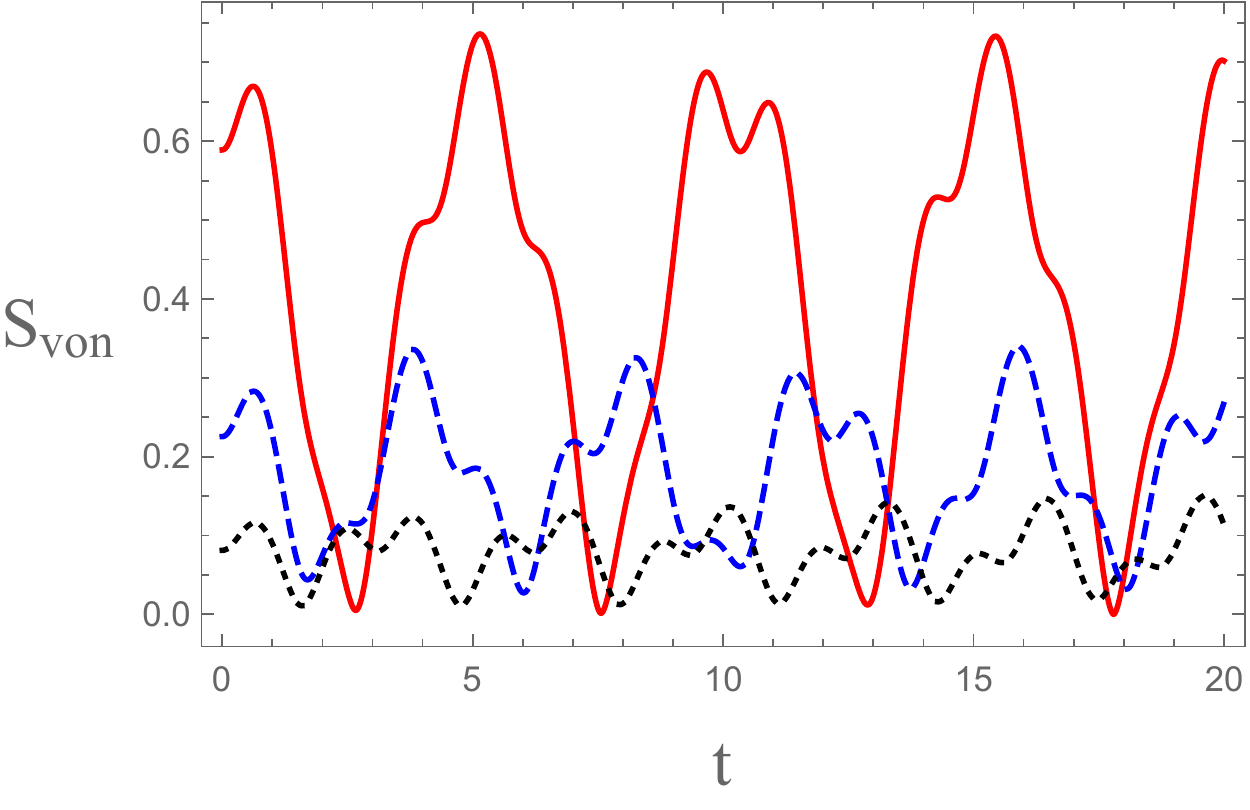} \hspace{0.2cm}
\includegraphics[height=5.0cm]{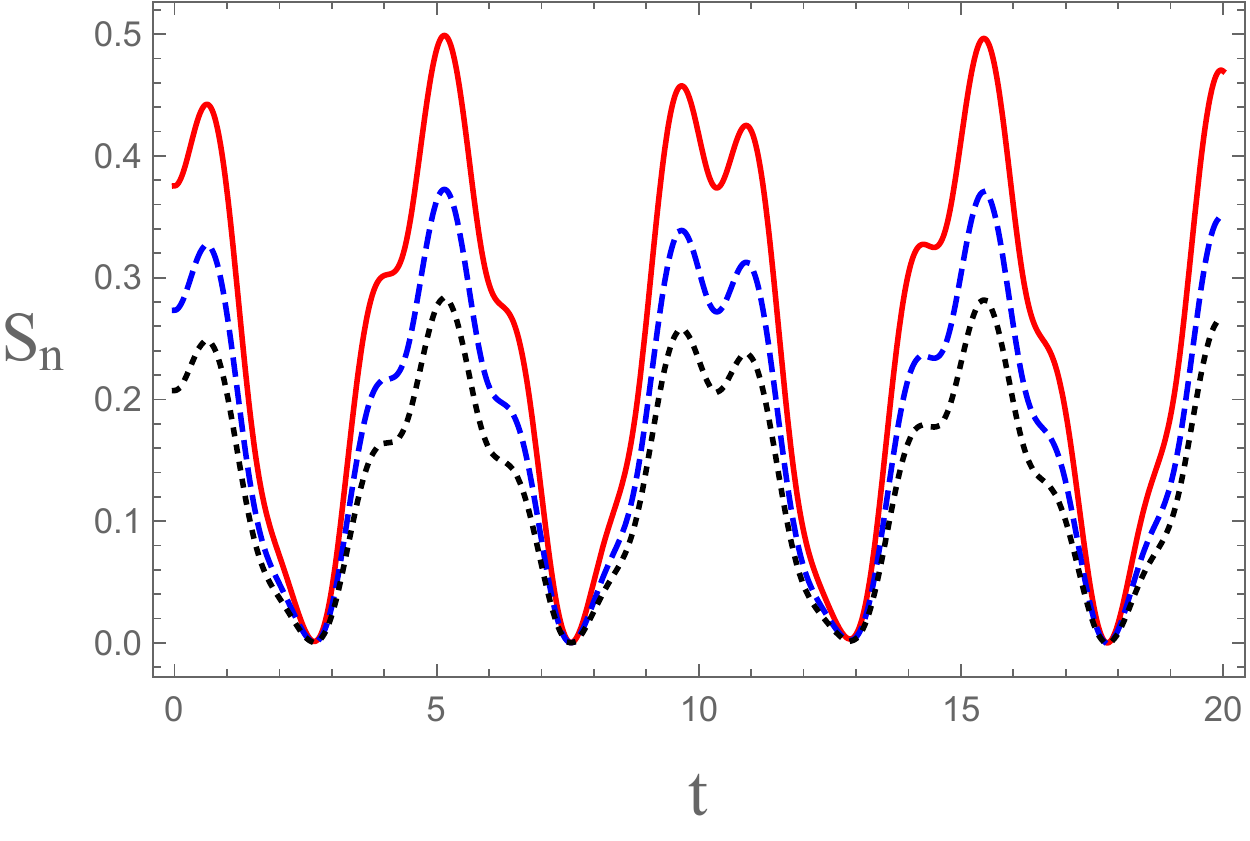}
\includegraphics[height=5.0cm]{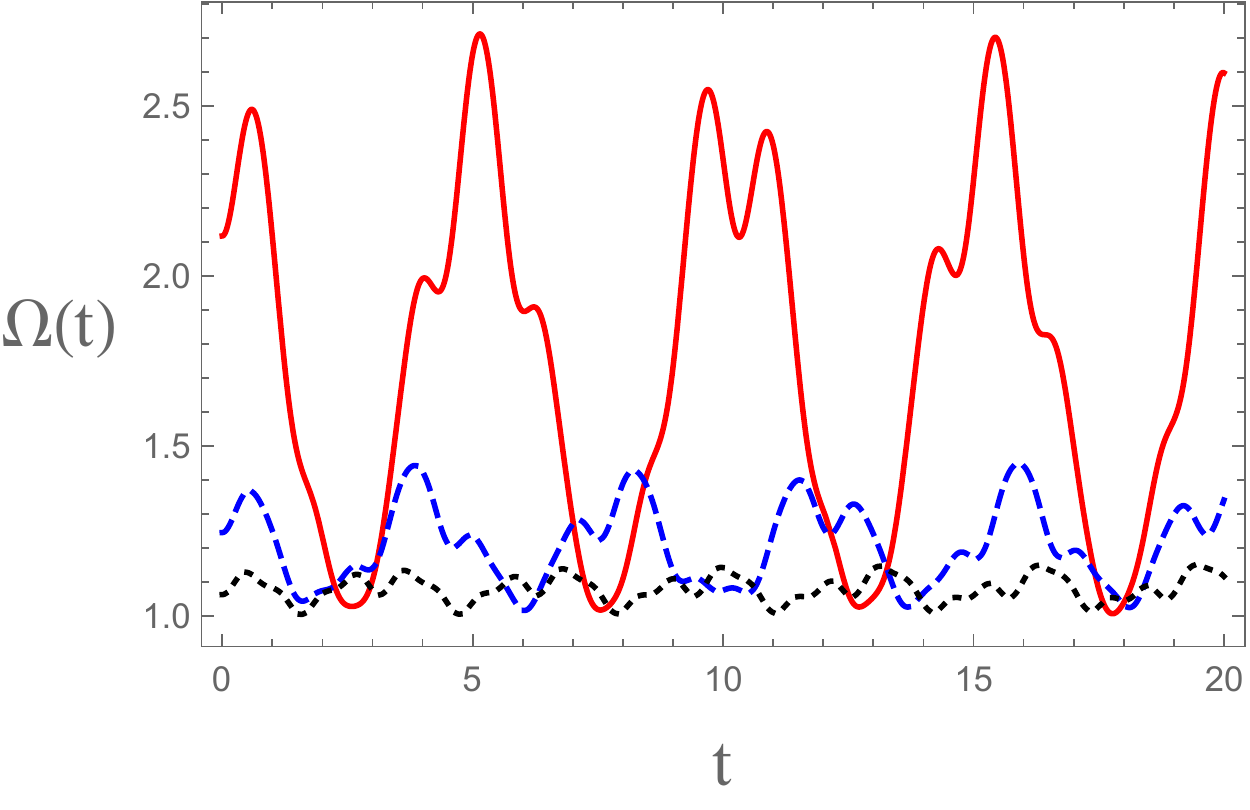}

\caption[fig3]{(Color online) The time-dependence of von Neumann entropy (Fig. 3(a)),  R\'{e}nyi entropy $S_n$ (Fig. 3(b)),  and 
uncertainty $\Omega (t)$ (Fig. 3(c)) when $\omega_{1, i} = 1$, $\omega_{1, f} = 1.3$, $\omega_{2, i} = 1.5$, and $\omega_{2, f} = 1.8$ for various $J$ 
in the realistic quenched model. (a) Dynamics of von Neumann entropy for $J = 1.1$ (red solid line), 
$J = 0.9$ (blue dashed line), and $J = 0.6$ (black dotted line). (b)  Dynamics of  R\'{e}nyi entropy for $n=2$ (red solid line), $n=4$ (blue dashed line), and 
$n = 100$ (black dotted line). In this figure $J$ is fixed as $1.1$. (c) Dynamics of uncertainty for $J = 1.1$ (red solid line), 
$J = 0.9$ (blue dashed line), and $J = 0.6$ (black dotted line). }
\end{center}
\end{figure}
%%%%%%%%%%%%%%%%%%%%%%%%%%%%%%%%%%%%%%%%%%%%%%%%%%%%%%%%%%%

The final and more realistic model we consider is a quenched model. In this model we choose the original angular frequencies $\omega_j$ as  
\begin{eqnarray}
\label{model3-1}
\omega_1 (t) = \left\{                \begin{array}{cc}
                                               \omega_{1, i}  & \hspace{0.25cm} t = 0   \\
                                                \omega_{1, f}   & \hspace{0.5cm} t > 0
                                               \end{array}            \right.         \hspace{1.0cm}
\omega_2 (t) = \left\{                \begin{array}{cc}
                                               \omega_{2, i}  & \hspace{0.25cm} t = 0   \\
                                                \omega_{2, f}   & \hspace{0.25cm}  t > 0.
                                               \end{array}            \right.
\end{eqnarray}
In this case the rotation angle $\alpha$ is completely determined by Eq. (\ref{angle}) if $J$ is given. Also $\tilde{\omega}_{1,i}$,  $\tilde{\omega}_{1,f}$,
$\tilde{\omega}_{2,i}$, and  $\tilde{\omega}_{2,f}$ are completely determined by Eq. (\ref{freq-1}). The scale factor $b_j (t)$ become
\begin{eqnarray}
\label{model3-2}
b_j (t) = \sqrt{ \left(\frac{\tilde{\omega}_{j, f}^2 - \tilde{\omega}_{j, i}^2}{2 \tilde{\omega}_{j, f}^2}\right) \cos (2 \tilde{\omega}_{j, f} t) 
+ \left( \frac{\tilde{\omega}_{j, f}^2 + \tilde{\omega}_{j, i}^2}{2 \tilde{\omega}_{j, f}^2}\right)} \hspace{1.0cm} (j=1, 2).          
\end{eqnarray}
The time-dependence of the von Neumann entropy $S_{von}$, R\'{e}nyi entropy $S_n$, and uncertainty $\Omega(t)$ is plotted in Fig. 3 when 
$\omega_{1, i} = 1$, $\omega_{1, f} = 1.3$, $\omega_{2, i} = 1.5$, and $\omega_{2, f} = 1.8$ with varying $J$ (Fig. 3(a), Fig. 3(c)) or $n$ (Fig. 3(b)).
In Fig. 3(a)  the von Neumann entropy is plotted for $J=1.1$ (red solid line), $J = 0.9$ (blue dashed line), and $J = 0.6$ (black dotted line).
Unlike the previous toy models the large $\alpha$ (or large $J$) does not guarantees higher entanglement in the full range of time in this realistic model. 
Another difference is a fact that the time-dependence of the von Neumann entropy exhibits a double oscillatory behavior. This is due to the fact that the 
trigonometric functions are involved in both $b_1 (t)$ and $b_2 (t)$. The time-dependence of the  R\'{e}nyi entropy is plotted  for $n=2$ (red solid line), $n=4$ (blue dashed line), and $n = 100$ (black dotted line). In this figure $J$ is fixed as $1.1$. It also exhibits a double oscillatory behavior. With increasing $n$ the 
R\'{e}nyi entropy decreases, and eventually approaches to $S_{\infty} = - \ln (1 - \xi)$. Most striking difference arises in the dynamics of the uncertainty 
$\Omega (t) = (2 \Delta x_1 \Delta p_1)^2$. This is plotted on Fig. 3(c) for $J = 1.1$ (red solid line), $J = 0.9$ (blue dashed line), and 
$J = 0.6$ (black dotted line). In the previous toy models large $\alpha$ yields small $\Omega (t)$ at large time region. However this does not hold 
in this realistic model. In this model large $J$ yields large $\Omega (t)$ in most region of time domain. The surprising fact is that 
$S_{von}$ and $\Omega$ exhibit similar pattern. We do not know whether or not this is universal property.  If so, one can use the uncertainty 
as a candidate of  entanglement measure after rescaling it appropriately. It also exhibits a double oscillatory behavior due to the scale factors $b_j (t)$. 

\section{Conclusions}
The dynamics of the entanglement and uncertainty relation is examined by solving the TDSE of the coupled harmonic oscillator system when the angular frequencies 
$\omega_j$ and coupling constant $J$ are arbitrarily time-dependent and two oscillators are in ground states initially. To show the dynamics pictorially we 
introduce two toy models and one realistic quenched model. While the dynamics can be conjectured by simple consideration in the toy models, the dynamics in the realistic quenched model is somewhat different from that in the toy models. In particular, the dynamics of entanglement exhibits similar behavior to dynamics of uncertainty parameter in the realistic quenched model. We do not know whether or not this is general feature. 

It is natural to ask how the dynamics of entanglement and uncertainty relation is changed in the excited states. This issue is examined in appendix A, where the two oscillators are 
in ground and first-excited states initially. In this case we fail to compute the entanglement analytically because we do not know how to derive the
eigenfunctions and the corresponding eigenvalues explicitly. However, the uncertainty relation is derived exactly in the appendix.

Another interesting issue related to the entanglement of the  coupled harmonic oscillators is multipartite entanglement. Consider the three coupled 
harmonic oscillator system, whose Hamiltonian is 
\begin{equation}
\label{con-8}
H = \frac{1}{2} (p_1^2 + p_2^2 + p_3^2) + \frac{1}{2} \left( \omega_1^2 (t) x_1^2 + \omega_2^2 (t) x_2^2 + \omega_3^2 (t) x_3^2 \right) 
- ( J_{12} (t) x_1 x_2 + J_{13} (t) x_1 x_3 + J_{23} (t) x_2 x_3).
\end{equation}
We conjecture that the TDSE of this system can be solved analytically. However, computation of the tripartite entanglement seems to be formidable task. 
First of all we do not know what kind entanglement measure can be computed. In qubit system we usually use the three tangle\cite{ckw} or 
$\pi$ tangle\cite{ou07-1} to measure the tripartite entanglement. However, it is not clear whether these tangles can be computed analytically in the coupled harmonic oscillator system or not. We hope to visit this issue in the future.

{\bf Acknowledgement}:
%On April 16, 2014 the ferry Sewol has sunk into the South Sea of Korea. Due to this disaster 304 people died and, 9 of them are still missing. We would like to dedicate this paper to all victims of this accident.
%This research was supported by the Basic Science Research Program through the National Research Foundation of Korea(NRF) funded by the Ministry of Education, Science and Technology(2011-0011971).
This work was supported by the Kyungnam University Foundation Grant, 2016.

\newpage 

\begin{appendix}{\centerline{\bf Appendix A}}

\setcounter{equation}{0}
\renewcommand{\theequation}{A.\arabic{equation}}

%%%%%%%%%%%%%%%%%%%%%%%%%%%%%%%%%%%%%%%%%%%%%%%%%%%%%%%%%
\begin{figure}[ht!]
\begin{center}
\includegraphics[height=5.0cm]{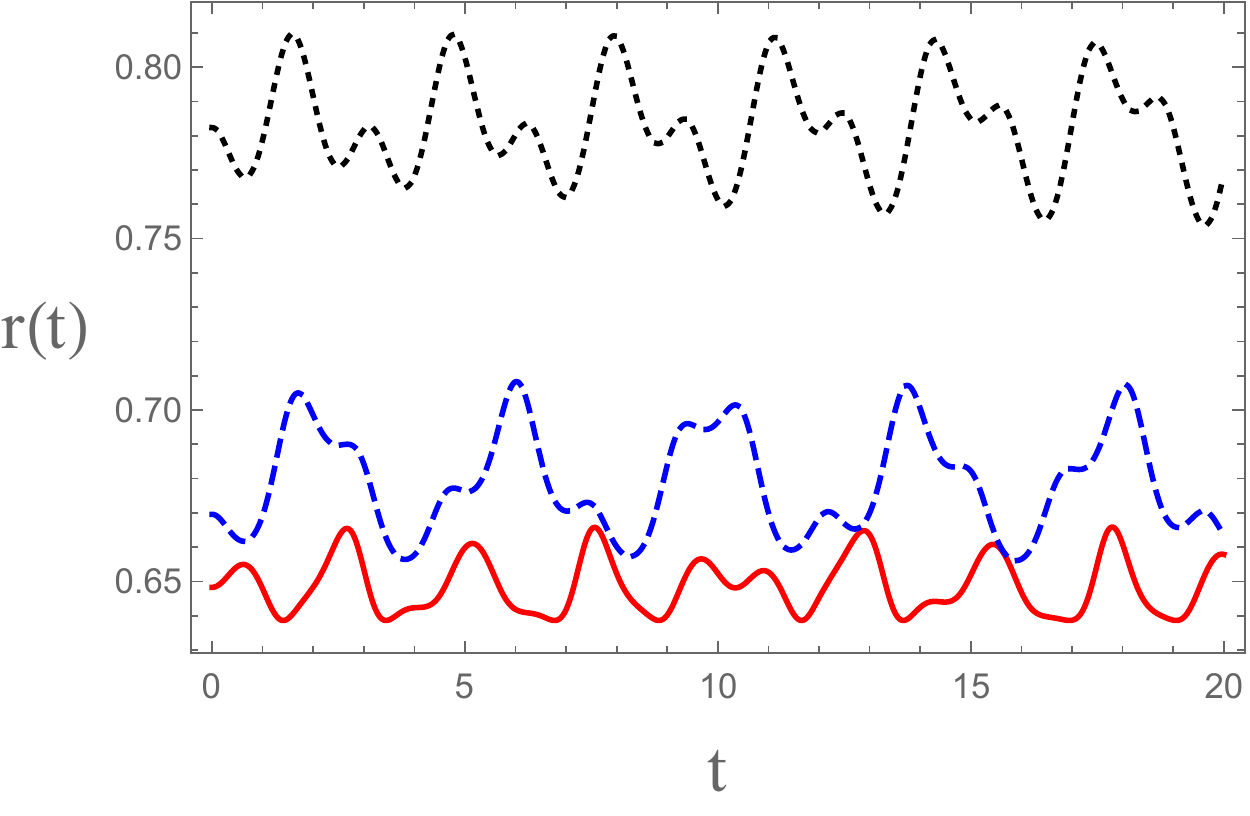} \hspace{0.2cm}
\includegraphics[height=5.0cm]{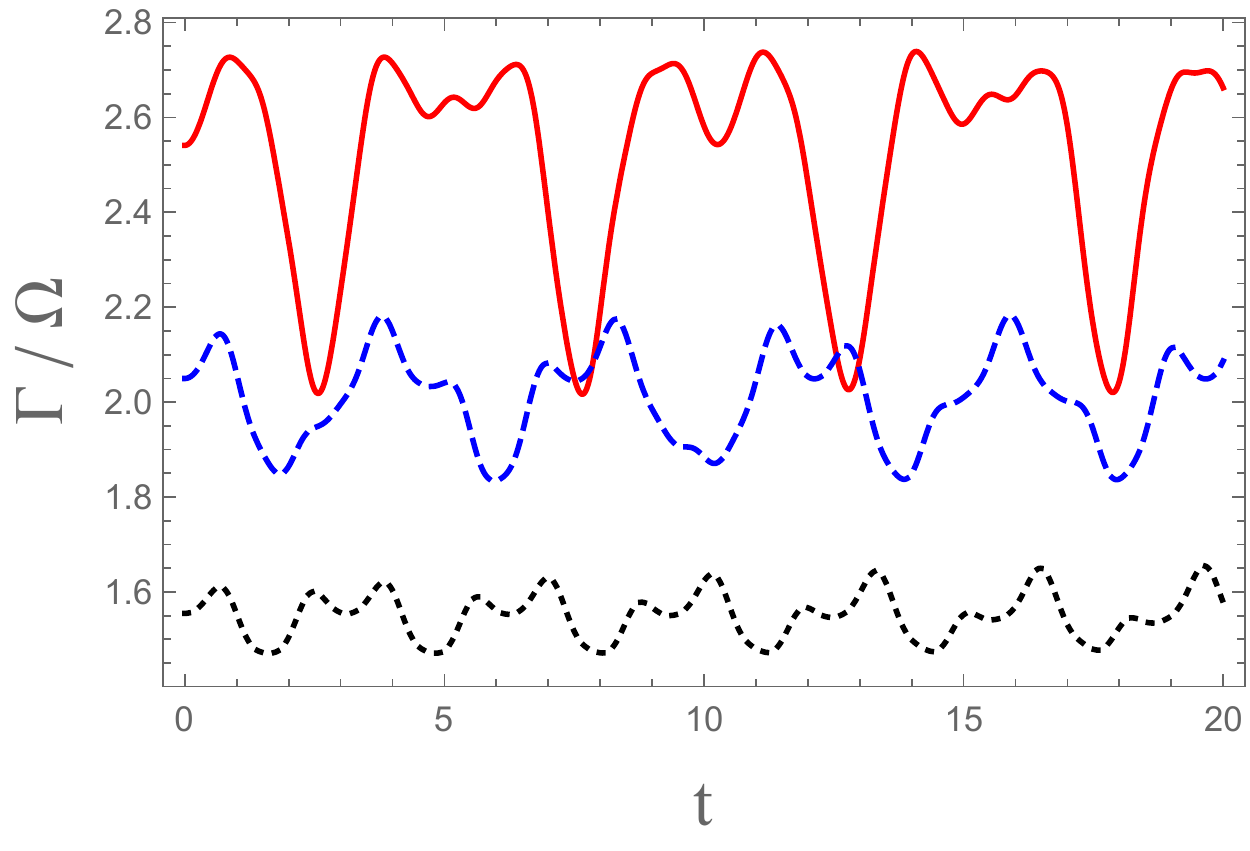}

\caption[fig4]{(Color online) The time-dependence of ratio for mixedness $r(t)$ (Fig. 4(a)) and uncertainties $\Gamma (t) / \Omega (t)$ ((Fig. 4(b)) 
between  $\rho_{(0,0)}^A$ and $\rho_{(0,1)}^A$  in the realistic quenched 
model. We choose $\omega_{1, i} = 1$, $\omega_{1, f} = 1.3$, $\omega_{2, i} = 1.5$, and $\omega_{2, f} = 1.8$ for  $J = 1.1$ (red solid line), 
$J = 0.9$ (blue dashed line), and $J = 0.6$ (black dotted line). Since $r(t) < 1$ in the full range of time, Fig. 4(a) indicates that  $\rho_{(0,1)}^A$ is more 
mixed state than $\rho_{(0,0)}^A$. It is of interest to note that  $\rho_{(0,1)}^A$  becomes more and more mixed compared to $\rho_{(0,0)}^A$ 
with increasing the coupling constant $J$. Fig. 4(b) shows that the uncertainty $\Delta x_1 \Delta p_1$ increases in  $\rho_{(0,1)}^A$  compared to that of 
 $\rho_{(0,0)}^A$. The increasing rate becomes larger with increasing the coupling constant $J$.
 }
\end{center}
\end{figure}
%%%%%%%%%%%%%%%%%%%%%%%%%%%%%%%%%%%%%%%%%%%%%%%%%%%%%%%%%%%

In this appendix we examine how to extend the main results of this paper to the excite states. 
If, for example, two oscillators are in ground and first-excited states initially, 
the reduced density matrix becomes
\begin{equation}
\label{con-1}
\rho_{(0, 1)}^A (x_1, x'_1 : t) = 2 \omega'_2 \rho_{(0, 0)}^A (x_1, x'_1 : t) \left[ \frac{\cos^2 \alpha}{2 D} + F_1 x_1^2 + F_1^* {x'_1}^2 + F_2 x_1 x'_1 \right]
\end{equation}
where $ \rho_{(0, 0)}^A (x_1, x'_1 : t)$ is given in Eq. (\ref{density-4}) and 
\begin{eqnarray}
\label{con-2}
&&F_1 = \frac{\sin^2 \alpha \cos^2 \alpha}{4 D^2} \bigg[ \left(\omega'_1 - \omega'_2 \right) 
\left\{\omega'_1 (1 + \sin^2 \alpha)  + \omega'_2 \cos^2 \alpha  \right\}                                    \\    \nonumber
&& \hspace{3.0cm}- 
\cos^2 \alpha \left( \frac{\dot{b_1}}{b_1} - \frac{\dot{b_2}}{b_2} \right)^2 - 2 i \omega'_1 \left( \frac{\dot{b_1}}{b_1} - \frac{\dot{b_2}}{b_2} \right) \bigg]                                                                                                                                     \\   \nonumber
&& F_2 = \frac{1}{D} (2 a_3 \cos^2 \alpha + \omega'_1 \sin^2 \alpha ).
\end{eqnarray}
The explicit expression of $a_3$ is given in Eq. (\ref{boso-2}).  Then one can show 
\begin{eqnarray}
\label{con-3}
&&\mbox{Tr} \left[\rho_{(0,1)}^A \right] \equiv \int dx \rho_{(0,1)}^A (x, x: t) = 1        \\     \nonumber
&&\mbox{Tr} \left[\left(\rho_{(0,1)}^A \right)^2\right] \equiv \int dx dx' \rho_{(0,1)}^A (x, x': t) \rho_{(0,1)}^A (x', x: t) = 
\mbox{Tr} \left[\left(\rho_{(0,0)}^A \right)^2\right] r(t)
\end{eqnarray}
where $r(t)$ is ratio of mixedness between $\rho_{(0,0)}^A$ and $\rho_{(0,1)}^A$ and its explicit expression is 
\begin{eqnarray}
\label{con-4}
&&r(t) = 4 {\omega'_2}^2  
 \Bigg[ \frac{\cos^4 \alpha}{4 D^2} + \frac{\cos^2 \alpha}{4 D a_1 (a_1 + 2 a_3)} 
\left\{ (F_1 + F_1^*) a_1 + (F_1 + F_1^* + F_2) a_3 \right\}                                                                                     \\   \nonumber
&&                                                                    
+ \frac{1}{16 a_1^2 (a_1 + 2 a_3)^2} \bigg[ a_1^2 \left\{ (F_1 + F_1^*)^2 + 4 |F_1|^2 + F_2^2 \right\}   
+ a_3^2 \left\{ 3 (F_1 + F_1^*)^2 + 3 F_2 (2 F_1 + 2 F_1^* + F_2) \right\}                                                                   \\    \nonumber
&&  \hspace{4.0cm}
+ 2 a_1 a_3 \left\{ (F_1 + F_1^*)^2 + 4 |F_1|^2 + F_2 (3 F_1 + 3 F_1^* + F_2) \right\} \bigg]   \Bigg].                                  
\end{eqnarray}
We expect that the entanglement between ground and first-excited harmonic oscillators is very small compared to that between two ground state harmonic oscillators. However, it is difficult to show this explicitly because the analytic derivation of eigenvalues and eigenfunctions for $ \rho_{(0,1)}^A (x, x': t)$ 
does not seem to be simple matter, at least for us. We hope to discuss the dynamics of entanglement for general excited $(m,n)$ state in the future.  
The time-dependence of the uncertainty $\Delta x_1 \Delta p_1$ for  $\rho_{(0,1)}^A$ can be computed analytically. The Wigner function $W_{(0,1)} (x_1, p_1, t)$ 
for this state becomes 
\begin{equation}
\label{con-5}
W_{(0,1)} (x_1, p_1, t) = W_{(0,0)} (x_1, p_1, t) \left[h_0 (t) + h_1 (t) x_1^2 + h_2 (t) p_1^2 + 2 h_3 (t) x_1 p_1 \right]
\end{equation}
where $W_{(0,0)} (x_1, p_1, t)$ is the Wigner function for $\psi_{0,0} (x_1, x_2, t)$  given in Eq. (\ref{wigner-5}) and
\begin{eqnarray}
\label{con-6}
&&h_0 (t) = \frac{\omega'_1 \omega'_2} {\bar{\eta}} \cos 2 \alpha                                   \\   \nonumber
&&h_1 (t) = \frac{2 \omega'_2 \sin^2 \alpha}{\bar{\eta}^2} \left\{ \left[ \omega'_1 \tilde{D} + \cos^2 \alpha \frac{\dot{b}_1}{b_1} 
\left( \frac{\dot{b}_1}{b_1} - \frac{\dot{b}_2}{b_2} \right) \right]^2 + \left[ \omega'_1 \frac{\dot{b}_2}{b_2} \sin^2 \alpha 
+ \omega'_2 \frac{\dot{b}_1}{b_1} \cos^2 \alpha \right]^2 \right\}                                    \\    \nonumber
&&h_2 (t) =  \frac{2 \omega'_2 \sin^2 \alpha}{\bar{\eta}^2} \left[ D^2 + \cos^4 \alpha \left( \frac{\dot{b}_1}{b_1} - \frac{\dot{b}_2}{b_2} \right)^2
                                                                                                                         \right]                       \\   \nonumber
&&h_3 (t) = \frac{2 \omega'_2 \sin^2 \alpha}{\bar{\eta}^2} \Bigg\{ \cos^2 \alpha \left(\frac{\dot{b}_1}{b_1} - \frac{\dot{b}_2}{b_2} \right) 
\left[ \omega'_1 \tilde{D} + \cos^2 \alpha \frac{\dot{b}_1}{b_1} \left( \frac{\dot{b}_1}{b_1} - \frac{\dot{b}_2}{b_2} \right) \right]   \\   \nonumber
&&  \hspace{9.0cm}+ 
D \left[ \omega'_1 \frac{\dot{b}_2}{b_2} \sin^2 \alpha + \omega'_2 \frac{\dot{b}_1}{b_1} \cos^2 \alpha \right] \Bigg\}.
\end{eqnarray}
Then, it is straightforward to show that the uncertainty relation for  $\rho_{(0,1)}^A$ becomes $(\Delta x_1 \Delta p_1)^2 = \Gamma (t) / 4$, where 
\begin{eqnarray}
\label{con-7}
&&\Gamma (t) = \left( \frac{\bar{\eta}}{\omega'_1 \omega'_2} \right)^2
\left[ \left(h_0 \alpha_2 + \frac{h_2}{2} \right) + \frac{3 \bar{\eta}}{2 \omega'_1 \omega'_2} \left(h_1 \alpha_2^2 + h_2 \alpha_3^2 + 2 h_3 \alpha_2 \alpha_3 \right) \right]                        \\    \nonumber
&&      \hspace{4.0cm} \times
\left[ \left(h_0 \alpha_1 + \frac{h_1}{2} \right) + \frac{3 \bar{\eta}}{2 \omega'_1 \omega'_2} \left(h_2 \alpha_1^2 + h_1 \alpha_3^2 + 2 h_3 \alpha_1 \alpha_3 \right) \right]
\end{eqnarray}
where $\alpha_j$ are defined in Eq. (\ref{wigner-6}).

The time-dependence of  the ratios $r (t)$ and $\Gamma (t) / \Omega (t)$  for the realistic quenched model is plotted in Fig. 4, where  
$\omega_{1, i} = 1$, $\omega_{1, f} = 1.3$, $\omega_{2, i} = 1.5$, and $\omega_{2, f} = 1.8$ are chosen. The red solid, blue dashed, and black dotted lines
correspond to $J=1.1$, $J=0.9$, and $J=0.6$ respectively. The fact $r(t) < 1$ in the full range of time indicates that  $\rho_{(0,1)}^A$ is more 
mixed  than $\rho_{(0,0)}^A$. It is of interest to note that  $\rho_{(0,1)}^A$  becomes more mixed compared to  $\rho_{(0,0)}^A$  
with increasing the coupling constant $J$. Fig. 4(b) indicates that  the uncertainty $\Delta x_1 \Delta p_1$ increases in  $\rho_{(0,1)}^A$  compared to that of 
 $\rho_{(0,0)}^A$. The increasing rate becomes larger with increasing the coupling constant $J$.

\end{appendix}

\end{document}